\documentclass[preprint2]{aastex}
\def\kms{\mbox{km~s$^{-1}$}}
\def\kpc{\mbox{kpc}}
\def\kpch{\mbox{$h^{-1}$ kpc}}

\def\LCDM{\mbox{$\Lambda$ CDM}}

\def\Mpc{\mbox{Mpc}}

\def\Mpch{\mbox{$h^{-1}$ Mpc}}
\def\Mvir{\mbox{M$_{\rm vir}$}}
\def\Msun{\mbox{M$_\odot$}}
\def\Msunh{\mbox{$h^{-1}\ M_\odot$}}

\slugcomment{draft ApJ, 07/01/03}

\shorttitle{SDSS satellites}
\shortauthors{Prada et al.}

\begin{document}

\title{Observing the dark matter density profile of isolated galaxies}

\author{Francisco Prada$^{1,2,3}$, Mayrita Vitvitska$^4$, Anatoly
Klypin$^4$, Jon A. Holtzman$^4$, David J. Schlegel$^5$, Eva K. Grebel$^1$,
H.-W. Rix$^1$, J. Brinkmann$^6$, T.A. McKay$^7$, \and I. Csabai$^{8,9}$}

\affil{$^{1}$Max-Planck-Institut f\"{u}r Astronomie, K\"{o}nigstuhl 17,
D-69117 Heidelberg, Germany}
\affil{$^{2}$Centro Astron\'omico Hispano-Alem\'an, Apdo 511, E-04080
Almer\'{\i}a, Spain}
\affil{$^{3}$Current address: Instituto de Astrof\'{\i}sica de Canarias, 
E-38205 Tenerife and The Isaac Newton Group of Telescopes, Apdo 321,
E-38700 La Palma, Spain}
\affil{$^4$Astronomy Department, New Mexico State University, Box 30001,
Department 4500, Las Cruces, NM 88003, USA}
\affil{$^5$University Observatory, Peyton Hall, Princeton, NJ 08544-1001, USA}
\affil{$^6$Apache Point Observatory, P.O. Box 59, Sunspot, NM 88349, USA}
\affil{$^7$Department of Physics, University of Michigan, 500 East University 
Avenue, Ann Arbor, MI 48109, USA}
\affil{$^8$Department of Physics, E\"{o}tv\"{o}s University,
Budapest, Pf.\ 32, H-1518, Hungary}
\affil{$^9$Department of Physics and Astronomy, The Johns Hopkins
University, 3701 San Martin Drive, Baltimore, MD~21218, USA}


\begin{abstract}

Using the Sloan Digital Sky Survey (SDSS), we probe the halo mass distribution
by studying the velocities of satellites orbiting isolated galaxies. In a 
subsample that covers 2500 sq.~degrees on the sky, we detect about 3000 
satellites with absolute blue magnitudes going down to $M_B = -14$; most of 
the satellites have $M_B=-16$ to $-18$, comparable to the magnitudes of 
M32 and 
the Magellanic Clouds. After a careful, model-independent removal of 
interlopers, we find that the line-of-sight velocity dispersion of satellites 
declines with  distance to the primary. For an $L_*$ galaxy the r.m.s. 
line-of-sight velocity changes 
from $\approx$$120~\kms$ at 20~\kpc~ to $\approx$$60~\kms$ at 350~\kpc. This 
decline agrees remarkably well with theoretical 
expectations, as all modern cosmological models predict that the density of 
dark matter in the peripheral parts of galaxies declines 
as $\rho_{\rm DM}\propto r^{-3}$. Thus, for the first time we find direct 
observational evidence 
of the density decline predicted by cosmological models; we also note that 
this result contradicts alternative theories of gravity such as MOND. We also 
find that the velocity dispersion of satellites within 100~\kpc~ scales with 
the absolute magnitude of the central galaxy as $\sigma \propto L^{0.3}$; 
this is very close to the Tully--Fisher relation for normal spiral galaxies.

\end{abstract}


\section{Introduction}

 Measuring the distribution of mass around galaxies at large radii
($\geq 100~\kpc$) provides a critical test for cosmological models. By
studying the mass distribution at these distances, we directly address
one of the most intriguing questions of modern cosmology -- the nature
of dark matter. Specifically, we wish to determine the extent of galaxy
dark matter halos and their density profile. While the inner parts of
galaxy halos have density distributions that yield approximately flat
rotation curves, our understanding of the profile at larger distances
is much poorer. The issue is critical because the density profile that
gives rise to a flat rotation curve ($\rho \propto r^{-2}$) is different
from that predicted by cosmological models \citep[NFW][]{NFW}  at 
larger distances ($\rho \propto r^{-3}$).

The challenge in measuring the mass distribution at large radii 
arises from the difficulty in finding a visible tracer to
probe the mass.  Historically, neutral hydrogen (HI) has
been used to study the outer parts of rotation curves 
\citep[e.g.,][]{RobertsRots,Bosma, vanAlbada}, since HI is detected 
well beyond the optical boundaries of spiral galaxies. The observations 
demonstrating that HI rotation velocities of
field spirals are roughly constant at large galactocentric distances
provides one of the primary pieces of evidence that galaxies are embedded
in massive dark matter halos \citep[e.g.,][]{Ostriker, Faber, Sancisi, Sofue}.
However, HI emission is detected only out to 30--50 kpc.  X-ray emission from
the diffuse hot gas in isolated elliptical galaxies also indicates
the presence of massive dark matter halos \citep[e.g.,][]{Buote94, Romanowsky,
Buote02}, but, just like HI in spiral galaxies, the X-ray emission does not
extend far enough to distinguish isothermal from cosmological $\rho\propto
r^{-3}$ profiles \citep{Buote02}. Strong gravitational lensing places
an important upper limit on the amount of dark matter but only in the inner
(few tens of \kpc ) regions of galaxies 
\citep[e.g.,][]{Keeton98,Keeton, Maller}.

Weak gravitational lensing provides a more promising method of studying
the outer parts of galaxies. Individual field galaxies produce a small
distortion of background galaxies, and this distortion can be used to
measure the galaxy--mass correlation function. Unfortunately, the existing
data \citep[e.g.,][]{Tyson, Brainerd, Hoeskstra, Fischer, Smith} do
not distinguish between a singular isothermal 
sphere \citep[SIS;][]{Schneider}
and a cosmological NFW profile: both models provide good fits
to the galaxy--mass correlation function in the outer regions where the
error bars are large.  Moreover, the galaxy--mass correlation function can be
affected by neighbors of the lens galaxy \citep[e.g.,][]{Smith, GuzikSeljak}. Once
the contamination from the neighboring galaxies is taken into account,
as done by, for example,  \citet[][]{Fischer, Hoeskstra}, \citet{McKay01} place
only a lower limit on the size of the halo.

Velocities of satellites of galaxies provide another way
to probe the mass distribution at large radii. These have been used
to constrain the mass of the Milky Way \citep{Zaritsky89, Lin95,
Kochanek96,EvansWilkinson2000} and M31 \citep{Evans}. However, because the
number of detectable satellites around galaxies outside of the Local Group
is small, one needs to study many galaxies to accumulate enough statistics
to study the profiles of other galaxies.  As a result, observational
efforts to study the dynamics of satellites have been somewhat limited
\citep{Erickson, Zaritsky93, ZaritskyWhite, Zaritsky97, McKay02}. Important 
early results
were obtained by \citet{Zaritsky93}, \citet{ZaritskyWhite} and
\citet{Zaritsky97}, who compiled
and studied a sample of about 100 satellites of nearby isolated spiral
galaxies with an average of 1--2 satellites per primary galaxy. It was
found that the line-of-sight velocity dispersion of the satellites does 
not decline with the projected distance to the primary galaxy. This result 
has been generally considered a strong argument for the presence of dark 
matter at large distances ($\sim$200--400 kpc) around galaxies. 
\citet{ZaritskyWhite} also found that the satellite velocity
dispersion does not correlate with the luminosity of the primary galaxy.

\citet{McKay02} used the SDSS to select and study a much larger sample of
1225 satellites. The mean number of satellites they found around each host 
was only two. The new analysis confirmed that the velocity dispersion
of satellites does not decline with distance from the primary, as observed
by \citet{Zaritsky93, ZaritskyWhite, Zaritsky97}.  This implies that
halos of isolated galaxies extend to distances of several hundred \kpc.
In contradiction with the earlier results, however, \citet{McKay02}
find that the velocity dispersion of satellites, $\sigma$, increases with
the luminosity, $L$, of the primary galaxy as $\sigma \propto L^{0.5}$ and
claim that this is consistent with the predictions of the semi-analytical models of galaxy
formation of \citet{Kauffmann} when the sample of primary galaxies in
the models is defined as it is in the observational analysis.

Cosmological models made definite predictions for the mass profile of
dark matter halos at large distances. Although SIS halo
models with density $\rho\propto r^{-2}$ and $\sigma = {\rm constant}$ 
are often {\it assumed}, there is little justification for
using SIS at these distances.  The main {\it motivations} for using SIS
are simplicity and extrapolation of flat rotation curves to
much larger radii.  However, during the last two decades of intensive
numerical modeling of galaxy formation, not a single model has produced
an SIS.  Every cosmological model studied so far (and there have been
plenty) has $\rho \propto r^{-3}$ at large radii. The slope does not
depend on the mean density of matter: CDM with
$\Omega_{\rm matter}=1$  \citep{Ghigna} and \LCDM\ with 
$\Omega_{\rm matter}=0.3 $ \citep{Klypin01}
have the same slope.  It does not depend on the nature of the dark
matter: warm dark matter \citep{Avila-Reese}, self-interacting dark
matter \citep{Colin02}, and cold dark matter all make the same
prediction. The slope does not depend on the halo mass: halos ranging
from cluster masses to dwarf masses all have $\rho\propto r^{-3}$. In
other words, the declining velocity dispersion of satellites is not a test for
the parameters of cosmological models, but rather of the hierarchical
scenario itself.  It is ironic that the same argument -- the constant
velocity dispersion of satellites -- which just a few years ago provided
one of the strongest cases for the existence of dark matter, is now
an argument {\it against dark matter}. The only model that predicts a
constant velocity dispersion is  Modified Newtonian Dynamics (MOND).

 As previous authors have noted \citep{Zaritsky92, ZaritskyWhite}, a 
comprehensive measurement of the satellite velocity dispersion profile 
around isolated galaxies should ideally combine two main elements: 
(1) a large number of primary galaxy and satellite (candidates) to 
estimate the satellite kinematics in separate bins of primary mass 
(or luminosity) and projected separation, and (2) a model-independent, 
robust, and testable rejection of interloper galaxies -- dwarfs with 
large physical distances from the primary galaxies but small projected 
and velocity differences -- that are not bound to the primary. The
effect of the interlopers is to make the halo mass profile difficult to 
measure, 
and that may have led to systematic effects in the few observations that 
suggest a constant velocity dispersion at large radii. Understanding 
 their effect is crucial for any interpretation of the halo mass 
distribution. Results from weak lensing can be affected by the presence 
of nearby projected neighbors, leading to similar systematic 
problems \citep[see][]{Smith, GuzikSeljak}.

 In this paper we use the vast data base of SDSS to study the motion 
of satellites around a carefully selected sample of isolated nearby galaxies.
We also develop an improved, model-independent approach for interloper
rejection, which, together with state-of-the art cosmological $N$-body 
simulations of galaxies and their satellites, allows us to measure the 
dark matter halo profile.  

In Section 2, we describe the
observational data and the criteria for selecting our samples of primaries and
satellites. The numerical simulations are briefly described in Section 3,
where we also discuss the use of simulations to test prescriptions to remove
interlopers. Our observational results are presented in Section 4, and our conclusions 
are given in Section 5. 

\section{Observational Data: Selection of primaries and satellites}

The SDSS \citep{York, Stoughton}
is a survey that  images up to $10^4$ $deg^2$ of the the northern Galactic
cap in five bands $ugriz$ \citep{Fukugita, Pier, Hogg, Smith02}
using a drift-scanning mosaic CCD camera \citep{Gunn}, down to a limiting 
magnitude of $r'\sim23$ \citep{Lupton}.  Approximately 
900,000 galaxies down to $r'\sim17.7$ will be
targeted for spectroscopic follow-up using two fiber-fed spectrographs
on the same telescope \citep{Strauss, Blanton02}.  

\begin{figure}[tb!]
\plotone{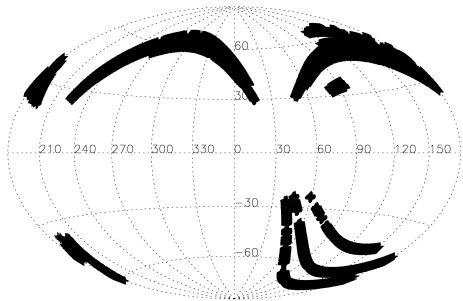}
\caption{Sky coverage for the SDSS dataset used in this work. The
total area covered by our data is $\sim$2500~deg$^2$.}
\label{fig:fig1}
\end{figure}

For the current analysis, we use data on galaxies for which spectra were
obtained before August 2002. We use ``survey quality'' redshifts derived from
these spectra by D. Schlegel. 
The recessional velocity errors are always less than 20~\kms~. Our SDSS 
redshift
sample consists of 254,073 galaxies distributed in several strips on
the sky; the total area covered by our data is about 2,500~deg$^2$.
Photometric parameters were taken from the PHOTO measurement 
\citep{Lupton} of the brightest object within 3 arcsec
that was loaded into the SDSS collaboration database in August 2002
(but see also discussion below); magnitudes were corrected for
foreground extinction using the \citep{Schlegel98} values loaded in 
the SDSS database. The
sky coverage of the resulting SDSS imaging and spectroscopic data
is shown in Figure~\ref{fig:fig1}.

We supplemented the SDSS data with data from the RC3 catalog 
\citep{deVacouleurs} for two
reasons. First, as we are identifying isolated galaxies (see below), we want
to avoid choosing galaxies that have bright neighbors outside the area
covered by the SDSS at the time we made our sample; the RC3 provides
a catalog of galaxies over the whole sky to make this possible. 
While it is true that the RC3 catalog is not complete, especially at
larger distances, missing companions in the SDSS data are much more likely
for the nearest galaxies for which the area on the sky
corresponding to our isolation criterion is large.  For these, the RC3 catalog
should be nearly complete.  Another, possibly more
important, reason for using the RC3 is that the version of PHOTO that
we were using did not provide accurate photometry for large and bright
galaxies owing to problems with the deblending algorithm.  As a result,
we used RC3 $B_T$ magnitudes (or photographical magnitudes if $B_T$ was
not available), corrected for foreground reddening,
for all of the SDSS galaxies for which we found a match in the RC3
catalog (within 0.5 arcmin on the sky and 100 km s$^{-1}$ in velocity); for
all of the other galaxies (which are the smaller galaxies for which
the SDSS photometry is accurate), we transformed the SDSS $g$
magnitude to $B$ magnitudes using the relation given by \citet{Yasuda}.

Our main galaxy list was selected from the full redshift sample by 
taking all galaxies with $500 < cz < 60,000~\kms$ and apparent magnitude 
$r'< 17.7$; the total number of selected galaxies is 206,352.  Distances
are derived assuming a smooth Hubble flow with $h=0.7$ 
($H_0 =$ 100 $h^{-1}$ km s$^{-1}$ Mpc$^{-1}$); the heliocentric
velocities were converted to the Local Group Standard of Rest before computing
distances.

For the study of satellite dynamics, we define primary galaxies to be
those with absolute blue magnitude brighter than $M_B=-19.0$. We put
no restriction on Hubble type. We use
three different isolation and selection criteria to define three
different samples of  \textit{isolated}
primaries. To be considered as isolated, a galaxy must have no other galaxies
within a magnitude difference $\Delta M$, projected separation $\Delta R$,
and velocity separation $\Delta V$. Satellites are defined as all objects
within a projected distance $\delta r$ and velocity difference $\delta v$
and being at least $\delta m$ magnitudes fainter than the primary.
Table~\ref{tab:table1} gives the parameters used to define each
sample, and the number and  median
distance of primaries, and the number of satellites. 

 The first two samples have identical isolation criteria, but Sample 2
is defined for primaries out to a larger distance, hence only bright
satellites can be seen for the more distant galaxies. The shallower sample
has more satellites per primary, but the more distant sample has an overall
larger number of total satellites.
The third sample mimics conditions used by \citet{McKay02}, in which the
isolation criterion is significantly relaxed.  It has a very large
search radius (almost 3~Mpc), but the luminosity of the neighbors can
be quite large: half that of the luminosity of the primary.  In other words,
this prescription can pick a bright member of a group of galaxies and
erroneously treat it as an ``isolated'' galaxy. As discussed below we obtain
similar results for all of the samples, suggesting that our results are 
statistically significant and do not depend on the exact isolation criterion.
Our main results are based on the first two samples.

\begin{deluxetable}{lrrrrrr} 
\tablecolumns{5} 
\tablewidth{0pc} 
\tablecaption{Selection and isolation criteria for SDSS samples} 
\tablehead{ \colhead{\small
Parameter} & \colhead{\small Sample 1} & \colhead{\small Sample 2} &
\colhead{\small Sample 3} }
\startdata Maximum depth of the sample (\kms) & 10,000 & 60,000 & 60,000 \\
Constraints on bright neighbors: \\
\quad Magnitude difference, $\Delta M$ &2.0 & 2.0 & 0.75\\ 
\quad Minimum projected distance, $\Delta R$ (\kpch) & 500 & 500 & 2000\\ 
\quad Minimum velocity separation, $\Delta V$ (\kms) & 1000 & 1000 & 1000 \\ 
Constraints on satellites: \\
\quad Minimum magnitude difference, $\delta m$ & 2.0 & 2.0 & 1.5 \\
\quad Maximum projected distance to primary, $\delta r$ (\kpch) & 350 & 350 & 500\\ 
\quad Maximum velocity separation with primary, $\delta v$ (\kms) & 500 & 500 & 1000 \\ 
Number of isolated galaxies &1278 & 88603 & 26807 \\ 
Number of satellites & 453 & 1052 & 2734 \\ 
Statistics of isolated galaxies with at least one satellite: \\
\quad Number & 283 & 716 & 1107\\ 
\quad Mean distance (\kms)  & 7100 & 14700 & 23170 \\
\quad Mean distance (\kms) for $-19.5<M_B<-20.5$ & 7244 & 9785 & 11076 \\
\quad Mean distance (\kms) for $-20.5<M_B<-21.5$ & 7697 & 15917 & 20854 \\
Limiting magnitude $M_B$  & $-$16.4 & $-$18.0 & $-$19.0 \\
\enddata
\label{tab:table1}
\end{deluxetable}

\begin{figure}[tb!]
\plotone{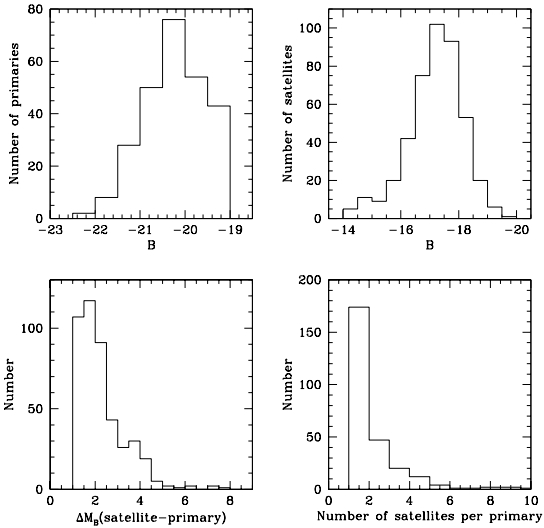}
\caption{Statistics of primary galaxies and their satellites in a 
subsample with velocities less than $10,000~\kms$.  The  top left panel
shows the distribution of absolute blue magnitudes, $M_B$, of isolated
primaries that have at least one satellite; the top right panel shows
the brightnesses of the satellites. The magnitude differences between the
primaries and the satellites is presented in the bottom left panel. The
number of satellites around each host is shown in the bottom right panel.}
\label{fig:fig2}
\end{figure}

Figure~\ref{fig:fig2} shows distributions of primary and satellite
brightnesses, brightness ratios, and number of satellites per primary
for Sample 1. Considering only primaries for which there is at least one
satellite, we find on average two satellites per primary in our first sample,
which is similar 
to that found by \citet{Zaritsky97} and \citet{McKay02}. A visual 
morphology classification of our primaries in Sample 1 indicates
that about 30\% of the galaxies are ellipticals and S0s (E--SO). This
fraction of early-type galaxies is simiar as seen in the RC3 catalog and 
in the SDSS  \citep{Nakamura}.  A fraction of 
our primaries is close to the edges of the strips and therefore we do 
not cover their entire halos; for 84$\%$ of the primaries the whole volume 
is covered.

Given the limiting magnitude of the SDSS spectroscopic survey, we expect to find
satellites down to $M_B \sim -14$ for a primary with $cz=2000$
km s$^{-1}$ and down to $M_B \sim -17$ for a primary with $cz=7100$
km s$^{-1}$, the mean of our primary recessional velocities in Sample
1. These limits would yield about four satellites per primary for our nearest
primaries and 1--2 satellites for the further ones if we assume that
all galaxies have satellite luminosity functions similar to that of the
Local Group \citep[see][]{Pritchet, Grebel}. Samples 2 and 3 include
more distant primaries, and only brighter satellites for the more distant
galaxies (see Table~\ref{tab:table1} for limiting magnitudes).

The main physical properties of satellites and primaries, such as
their color distribution, spectral properties, and
their subdivision into early- and late-type galaxies
will be discussed in \citet{Vitvitska}.

\section{Numerical models}
\label{ref:SecModels}

 Cosmological simulations of galactic dark matter satellites in the
hierarchical model of structure formation predict a remarkably large
number of dark matter satellites orbiting around a Milky Way-size halo
\citep{Klypin99, Moore}. In
many respects the simulations are quite realistic; we use them to
develop a method for removing the effects of interlopers on the amplitude
and shape of the r.m.s. velocity profile of the satellites. Note that the
results of the simulations are not used in the reduction of the real
astronomical data.  

We use one of the simulations presented by \citet{Klypin01}.  The
simulation is done for the standard $\Lambda$ CDM cosmological model
with $\Omega_0=0.3$, $\Omega_{\Lambda}=0.7$, and $h=0.7$; the spectrum
normalization is $\sigma_8=0.9$.  The Adaptive-Refinement-Tree (ART)
code \citep[see][for details]{Klypin01} is used to run
the simulation. The simulation box was $25\ \Mpch$. The formal force
resolution was $100\ h^{-1}$ pc; the mass resolution is $1.2\times 10^6~\Msunh$.

The simulation has three galaxy-size halos. All halos are relaxed and
have cusps. The outer parts of the density profiles of the halos are well
approximated by an NFW profile.  The distance between two of the halos is
600~\kpc; we decided not to use this pair because the halos are not
isolated. We selected the third halo, which does not have a massive
companion within 3~\Mpc~ radius. This halo has a virial mass of
$1.5\times 10^{12}\ \Msunh$, a virial radius of 235~\kpch, and a
maximum circular velocity of 214~\kms.  There are 200 dark matter
satellites inside a 350~\kpch~ radius from the center of the halo. Most
of the satellites are very small with circular velocity of
10--15~$\kms$ and mass $5\times 10^7\ \Msunh$.

The simulations were done in such a way that only regions around the
three large halos ($\sim$1/2 Mpc) have high resolution. The rest of
the computational volume was simulated with significantly lower
resolution. No halos outside  the regions of high resolution are
used in our analysis.



\subsection{Satellites in  cosmological simulations: removal of interlopers}
\label{ref:SecRemove}
\label{s21}
In any sample of isolated bright galaxies and their satellites there
is a fraction of galaxies that  are probably not related to the host, but
which are included in the sample because of projection effects. We
call these objects interlopers. The interlopers tend to have large
velocity differences and large projected separations. The fraction of
interlopers in the samples of \citet{ZaritskyWhite} and \citet{Zaritsky97} was
$15\%$. We expect a similar level of contamination in our
observational samples. If the interlopers are not removed, the r.m.s.
velocity of satellites, and the inferred mass of the primaries, 
will be overestimated. We use the simulations described in the last section
to test prescriptions for removing the interlopers.

\begin{figure}[tb!]
\plotone{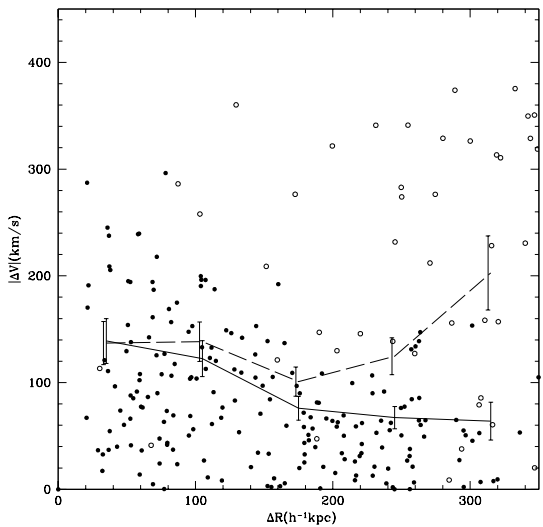}
\caption{The absolute value of the line-of-sight velocity difference 
between satellites and the central halo as a function of projected distance
to a simulated dark matter halo in the standard \LCDM~ model. Filled circles
show (true) satellites with 3D distance less than 350~\kpch~ from the
central halo with mass $M_{\rm vir} =1.5\times 10^{12}\Msunh$.  Open
circles show a realization of interlopers, which were generated by
randomly placing objects  inside a cylinder of radius
350~\kpch~ and length  1000~\kms. 
The number of interlopers 
is defined by the  luminosity function of galaxies in the SDSS. 
The dashed curve gives the r.m.s. velocity in the combined sample.
The full curve shows the r.m.s. velocity of true satellites.  }
\label{fig:fig3}
\end{figure}
 
To understand how interlopers might affect the results we introduce
them into the simulation. We have to do this because the $N$-body
simulation resolves only the region very close to the central halo; no
interlopers are present in the simulation.  
To make a sample that realistically mimics the observational data
we start by making
an estimate of the number of expected interlopers in the observed
sample.  Using the SDSS field luminosity function \citep{Blanton} we
find that 0.43 galaxies between $-18<M_B<-16$ are expected inside a
volume $R< 350~\kpch$ and with a velocity dispersion of
 $|\Delta V|<500~\kms$ at a distance of
$\sim$5000~\kms.  This is the estimated number of interlopers per primary, 
if galaxies are randomly distributed in space. Neglecting the large-scale
correlations of galaxies appears to be a reasonable assumption at this
point, because our primaries are selected to avoid clusters and groups 
of galaxies. The latter are responsible for most galaxy clustering on 
a scale of a few megaparsecs. Thus, the clustering of galaxies (with the 
exception of true satellites) around our primaries is  small and can be 
neglected. Using simulations this point should be quantified in
more detail in the future.

In the observed samples we find on average two satellites per primary.
Thus, we expect 0.215 interlopers for each true satellite. In our
numerical simulation we have 200 true satellites. This means that we
expect about 43 interlopers.  We randomly place 43 interlopers into a
cylinder of radius $R =350~\kpch$ and length $1000~\kms$.

Figure~\ref{fig:fig3} shows the resulting velocity differences
for both satellites and interlopers  as a
function of projected distance to the parent dark matter halo, as well as 
the r.m.s. 
velocities for the true satellites  and for the combined sample of the
satellites and interlopers.  Since the interlopers are
not physically associated with the primaries, their number depends
only on the volume.  Therefore, their effect on the r.m.s. velocity is
stronger at larger distances.  Neglecting interlopers clearly leads to
an apparent increase in the r.m.s. velocity with distance, while the
r.m.s. velocity of the true satellites actually decreases with
distance.

\begin{figure}[tb!]
\plotone{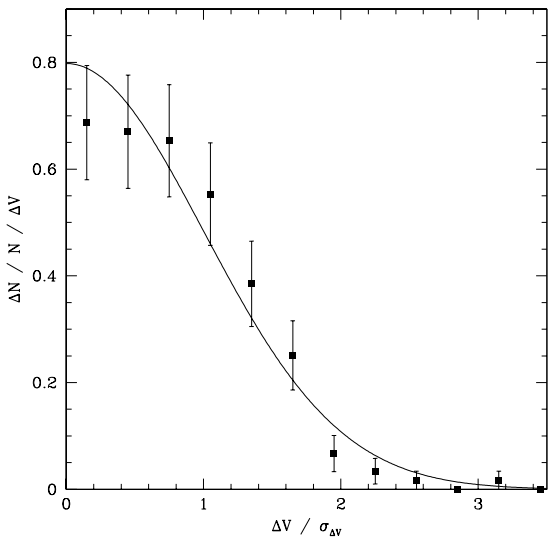}
\caption{The distribution of relative velocities of true satellites in
$N$-body simulations. The distribution is very close to  Gaussian
 (full curve). The distribution was obtained by finding the
r.m.s. velocities for individual bins and then  combining the
normalized velocities.}\label{fig:fig4}
\end{figure}

To reliably estimate the satellite velocity dispersion, one needs to
find a way to account for the interlopers.  We start by characterizing the
velocity distribution of true
satellites. Motivated by Figure \ref{fig:fig3}, we consider
a Gaussian velocity distribution, but allow the width of the distribution
to change with radius.
Figure~\ref{fig:fig4} shows the resulting distribution of normalized
(to the {r.m.s.} velocity in different radial bins)
velocities, and demonstrates that the distribution is
very close to  Gaussian. In reality, the distribution cannot
be completely Gaussian because satellites with very high
velocities will not be bound to the halo, and will escape.
Using simple
arguments based on the virial theorem, one naively expects that
satellites with velocities twice the r.m.s. velocity should not be
bound. Indeed, we find an indication of this: in
Figure~\ref{fig:fig4} the points are systematically below the
Gaussian fit for $\Delta V > 2\sigma$. However, it is difficult to use the
simple argument because we measure only one component of the velocity
and because the escape velocity changes with  distance. In any
case, the Gaussian fit provides a reasonably accurate approximation for
the distribution of the line-of-sight velocities of true satellites.

A simple prescription far removing the interlopers was used by
\citet{McKay02}. The velocity difference distribution of the entire
sample of neighbors (including objects at all projected distances) was
fitted by a Gaussian~+~constant function. The idea is that the
Gaussian describes the velocity distribution of satellites, and that  the
constant takes into account the distribution of interlopers; the width
of the Gaussian component gives a direct measurement of the r.m.s.
velocity of the satellites.  Our simulations show that the velocities
of the true satellites are indeed well approximated by the Gaussian
distribution, but that the width of the distribution (i.e., the
velocity dispersion) is a function of radius.  We find that the
\citet{McKay02} prescription has two problems that result from making
a single fit over all radii:

{\bf (1)} The level of interloper contamination (the constant in the fit)
does not depend on the distance to the primary. This is inaccurate
because of a purely geometrical factor: the volume and thus, the
number of interlopers, is larger for larger radii. Neglecting this
dependence produces a bias: the r.m.s. velocity is
overestimated at large radii and underestimated at small radii. 

{\bf (2)} There is no allowance for  dependence of the velocity dispersion
on projected distance. Because this dependence is in fact what we wish
to study, we do not want to impose the assumption of a constant velocity
dispersion on the prescription for the removal of interlopers.

In fact, if we apply the \citet{McKay02} prescription to the simulations,
we find that the recovered profile of the (interloper-removed) r.m.s.
velocity is flat; the prescription fails to recover the true dispersion profile
of the satellites. In this article, we apply an improved interloper 
prescription by making a separate ``Gauss~+~constant'' fit
for separate radial bins. In this case the increase of the number of
interlopers with the projected distance is taken into account (at least
approximately). This also allows for a radially dependent velocity 
dispersion of true satellites. In practice we use $100~\kpc$~ radial 
bins (70~\kpch~ in numerical simulations) and 100~\kms~ velocity bins. We do 
not include
satellites with projected distance less than $20~\kpc$. The binned data
are used to  minimize  the sum over all bins of normalized deviations
$(V_{i, \rm obs} -V_{i, \rm model})^2/N_i$, where $N_i$ is the number of
velocities in the  $i$th bin, and $V_{i, \rm obs}$ and $V_{i, \rm model}$ are the
observed and modeled velocities in the bin.

\begin{figure}[tb!]
\plotone{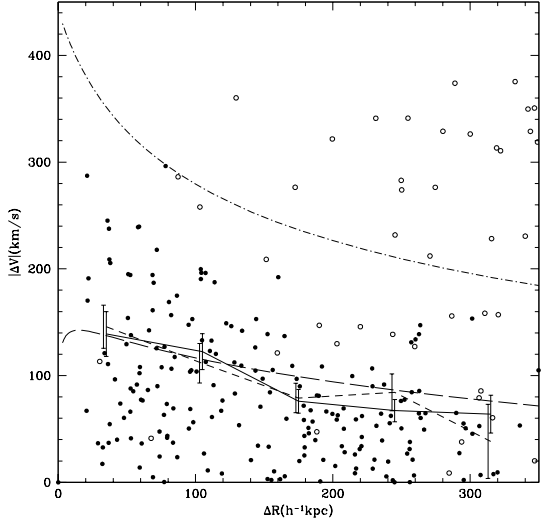}
\caption{The same as in Figure~\ref{fig:fig3}. The long dashed
   and dot-dashed curves show respectively the velocity dispersion and
   the escape velocity for the NFW profile with the same virial mass
   as the real dark matter halo. Most of the interlopers are above the escape
   velocity -- they cannot possibly belong to the dark matter halo. The r.m.s.
   velocity of the true satellites is plotted using the solid curve.
   This r.m.s. velocity of true satellites is accurately recovered
   (short-dashed curve) from the total sample of
   satellites~+~interlopers by making the Gaussian~+~constant fit for
   velocities in individual radial bins.}\label{fig:fig5}
\end{figure}

Figure~\ref{fig:fig5} shows the accuracy of the recovery of the
true r.m.s. velocity for the simulation; the short-dashed line shows the
recovered velocity while the solid line shows the true velocity
distribution of the satellites.  The procedure works remarkably
well. It recovers the value of the velocity dispersion and the number
of interlopers.  In the next section, we apply the same procedure to
the observational samples.

The theoretical prediction for the velocity dispersion profile for an NFW
profile with the same virial mass as that of the simulated dark matter halo is shown
by a long-dashed line.
In order to make this prediction for the r.m.s. line-of-sight
velocity, we use the NFW profile and assume that the satellites are in
equilibrium. We then use the Jeans equation to find the velocity
dispersion. The velocity anisotropy $\beta(r) \equiv
1-\sigma_{\perp}^2/2\sigma_r^2$, which is needed for the equation, was
taken from cosmological simulations \citep{Colin00,
Vitvitska02}. (Here $\sigma_{\perp}$ and $\sigma_r$ are the tangential and
radial r.m.s. velocities). The velocity anisotropy changes from very
small values $\beta \approx 0 $ close to the center of a halo to
$\beta\approx 0.5-0.6$ at the virial radius. We use this dependence of
$\beta$ on radius to make analytical estimates of the projected
velocity dispersion. Specifically, we integrate the r.m.s. velocities,
which we obtain from the Jeans equation, along a line of sight. The
integration is truncated at 1.5 virial radii. The velocity anisotropy
enters these calculations twice: through the Jeans equation and during
the integration along the line of sight. For a fixed mass profile $M(r)$
the effects of the velocity anisotropy are rather mild.  The
line-of-sight r.m.s. velocity in the realistic non-isotropic halo
declines with distance slightly more rapidly as compared with the isotropic
case ($\beta=0$).

\section{Results}

\subsection{Radial dependence of satellite velocity dispersion}
\label{s22}

\begin{figure}[tb!]
\plotone{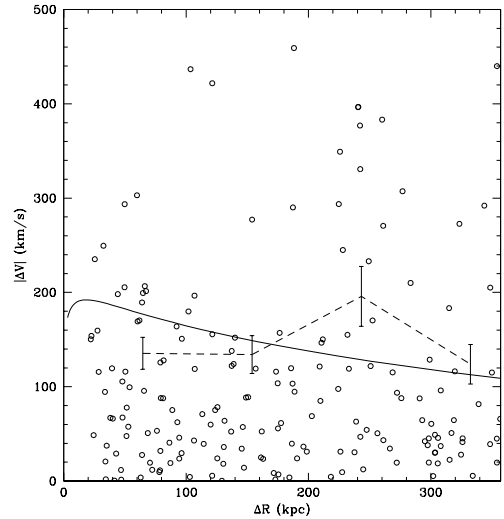}
\caption{The line-of-sight velocity differences of satellites in SDSS
   Sample 2 (open circles) for primaries in the range $-19.5 < M_B <
   -20.5$.
  The dashed curve shows the
  r.m.s. velocities for raw data with no removal of interlopers (except for
  the $\pm 500~\kms$ cut of the velocities). The r.m.s. velocity is
  basically flat; it does not depend on the distance to the
  primary. The full curve shows theoretical prediction for
  equilibrium NFW halo with mass $\Mvir = 5\times 10^{12}\ \Msun$. The
  decline in  velocities is not observed in the raw data.}
\label{fig:fig6}
\end{figure}

\begin{figure}[tb!]
\plotone{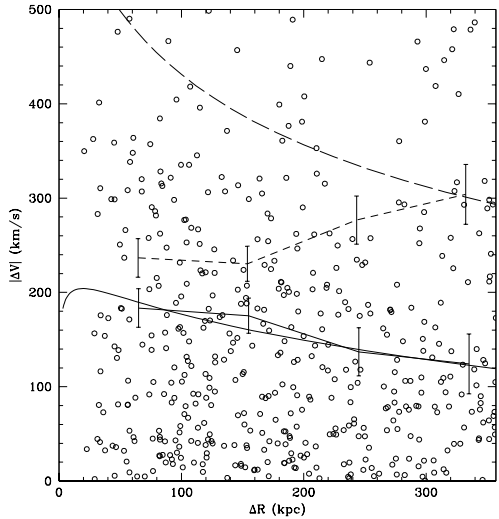}
\caption{The same as in Figure~\ref{fig:fig6}, but for Sample 3
    with primaries in the range $-20.5 < M_B < -21.5$. 
    The short dashed curve shows the r.m.s. velocities for
    raw data; the r.m.s. velocity increases slightly  with the distance.  
    The full curve with the error bars
    shows the r.m.s. velocity after removal of interlopers in each radial
    bin; the r.m.s. velocity is clearly declining with  distance. It
    is consistent with the NFW prediction (full curve) for halos with
    virial mass, $\Mvir =6\times 10^{12}\ \Msun$. 
    The long dashed curve 
    shows the escape velocity from the NFW halo of this mass. }
\label{fig:fig7}
\end{figure}
\begin{figure}[tb!]
\plotone{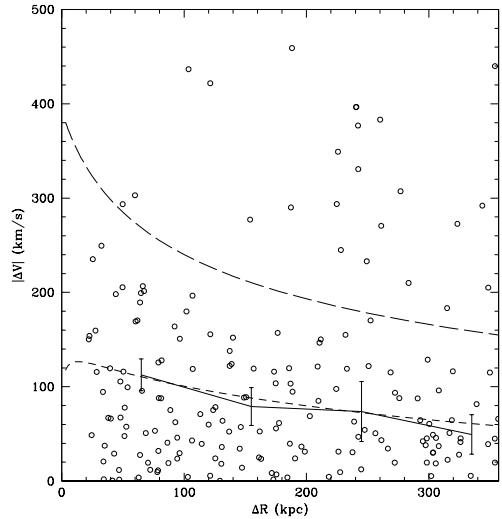}
\caption{
   The same as in Figure~\ref{fig:fig6}, but with the removal
   of the interlopers.  The full curve with the error bars shows the
   r.m.s. velocity after removal of interlopers. As in 
   Figure~\ref{fig:fig7}, the r.m.s. is clearly declining and
   consistent with the NFW profile (short-dashed curve) with $\Mvir
   =1.5\times 10^{12}\ \Msun$. The long dashed curve shows the escape
   velocity from the NFW halo of this mass. All satellites above the
   escape velocity curve are interlopers.}
\label{fig:fig8}
\end{figure}

Figure~\ref{fig:fig6} shows the relative line-of-sight velocities
for a subset of Sample 2; only companions of galaxies with $-19.5
< M_B < -20.5$ are shown.  Figure \ref{fig:fig7} shows a comparable
plot for Sample 3, but only for the brighter primaries, since the
statistics are poor for the fainter primaries.  The raw r.m.s.
velocity is shown by the dashed curve. Both figures show that the
r.m.s. velocity of the raw (uncorrected) data either does not change or
slightly increases with the distance from the primary.  A few years
ago this would have been a strong argument in favor of dark matter
in the outer parts of galaxies, but as discussed in the introduction,
a flat r.m.s. velocity curve extending to 300~\kpc~ strongly
contradicts any existing dark matter model.  Taken at face value, the
raw r.m.s. velocities presented in Figures~\ref{fig:fig6} and
~\ref{fig:fig7} contradict the dark matter scenario, and actually
favor MOND.

However, if we remove interlopers using the Gauss~+~constant technique
discussed in the previous section, as shown by the full lines with
error bars in Figures \ref{fig:fig7} and \ref{fig:fig8}, we
find that the velocity dispersion of satellites declines with
distance. We note that the results for Sample 1, which extends to
fainter satellites, are similar to those shown here for the other two
samples.  Using Monte Carlo simulations, we reject a hypothesis that
the r.m.s. velocity is constant at the 97\% level corresponding 
approximately to the
 3$\sigma$ level for Gaussian statistics. For the
simulations we combine Samples 2 and 3, which provide
independent measurements. We assume that deviations of the r.m.s.
velocities are Gaussian in each of four bins for each sample. The
average value of the r.m.s. velocity is defined by four random
points. The probability of having $\chi^2$ larger than the observed
$\chi^2$ and   declining curves in two independent realizations
is $2.9\times 10^{-2}$, based on 10,000 realizations. Most of the
constraints come from  Sample 2, where the first and the last
points deviate from the average by 1.5$\sigma$ and 1.8$\sigma$
correspondingly. Sample 3 alone is consistent with the constant r.m.s.
velocity.  Still, even in this case the velocity declines. When we
combine both samples, the constant r.m.s. velocity becomes very
improbable.

The number of estimated interlopers is quite modest. For
Sample 1, which has 132 satellites for primaries in the range $-19.5
<$M$_B$$< -20.5$, we estimate that 23 are interlopers: 17\% of the
sample. The numbers are similar for the other samples: 19\% 
in Sample 2 and 20\% in sample 3. As expected from geometrical
arguments, the number of interlopers increases with the radius. For
example, in Sample 3 the number of estimated interlopers is 16.7
for radii in the range 100--200~\kpc, and 54 for 200--360~\kpc,
which is roughly consistent with a
constant number of interlopers per unit volume.

The observed decline in velocity dispersion with radius closely
matches that predicted by an NFW dark matter density profile.  Figures \ref{fig:fig7} and
\ref{fig:fig8}  give the predicted NFW velocities for
halos of mass $6\times 10^{12}\ \Msun$ (for primaries with
$-20.5<M_B<-21.5$) and $M_{\rm vir}=1.5\times 10^{12}\ \Msun$ (for primaries
with $-19.5<M_B<-20.5$).

If we remove interlopers using the method of \citet{McKay02}, the mean
velocity dispersion for each magnitude bin is smaller than the raw value,
but we do not get a declining velocity curve. This explains the difference
between our conclusions and those of \citet{McKay02}, who do not find a
radial dependence on velocity dispersion; as discussed above, we feel
that our procedure of allowing for the radial dependence of interloper
contamination is better  motivated physically.

\subsection{Relation of satellite velocity dispersion with host luminosity}


A comparison of Figures~\ref{fig:fig7} and \ref{fig:fig8}
clearly indicates that the r.m.s. velocity increases with the
luminosity of the primary. This contradicts the earlier results of
\citet{Zaritsky97}. Because the NFW profile provides an accurate fit to
the observed velocity dispersion profile,
we can use it to estimate the virial mass of each group of galaxies. For galaxies
with $-19.5<M_B<-20.5$ we find $M_{\rm vir}=1.5\times 10^{12}\ \Msun$.  For the
brighter magnitudes, $-20.5<M_B<-21.5$, the virial mass is $M_{\rm vir}=6\times
10^{12}\ \Msun$. These values give us virial mass-to-light ratios
$M/L=100$ and $M/L=150$ for the dimmer and brighter galaxies
correspondingly. Thus, the $M/L$ ratio increases with the luminosity
roughly as $M/L\approx L^{0.5}$. We believe that this change in
$M/L$ with luminosity is real and does not reflect a change
in the stellar populations for both magnitude bins. A reason for
this is that we did not see a dependence of the fraction of 
E--S0 with luminosity for the primaries in Sample 1, at least 
from $-$19 to $-$22 blue absolute magnitudes. A similar
increase of $M/L$ with luminosity ($M/L\propto L^{0.4 \pm 0.2}$) has been seen 
from weak lensing studies by \citet{GuzikSeljak}. See, however, 
\citet{McKay01} who,  also from weak lensing, found no dependence of 
the $M/L$ on luminosity. 

\begin{figure}[tb!]
\plotone{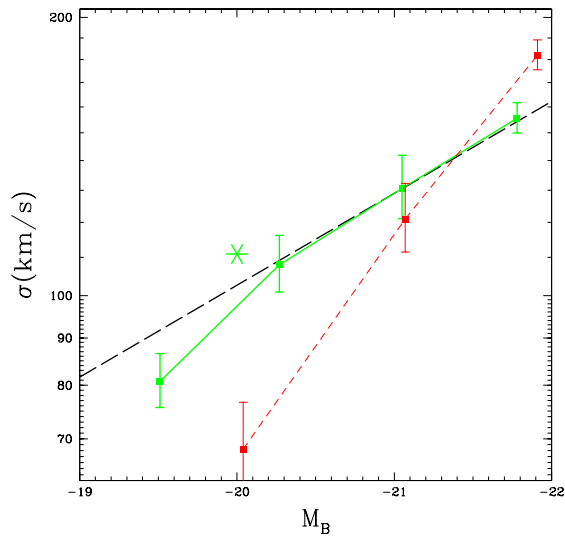}
\caption{
The dependence of satellite r.m.s. velocity on the absolute $B$ magnitude
of the primary galaxy. The solid line shows r.m.s. velocity inside 120~kpc.
The long-dashed line shows the Tully--Fisher relation of spiral galaxies 
\citep{Verheijen} scaled down by a factor of 1.6 to account for the 
difference in definitions of velocities.  The short-dashed line shows the 
r.m.s. velocity for satellites in the outer radial shell 250--350 kpc. For
these large radii the r.m.s. velocity is consistent with
$\sigma \propto L ^{0.5}$ (constant $M/L$). 
The asterisk shows the weak lensing results for $L_*$ 
galaxies \citep{Hoeskstra} 
}
%
\label{fig:fig9}
\end{figure}

Instead of trying to estimate masses of the galaxies (which is bound
to be model-dependent), we can study the more direct dependence of the
r.m.s. velocity on the luminosity of the primary galaxy. This can be
viewed as an analog of the Tully--Fisher relation.  Figure~\ref{fig:fig9}
presents the dependence of the satellite r.m.s. velocity, computed
within 120~kpc as a function of galaxy luminosity. We calculate the
r.m.s. velocity relatively close to the primary galaxy because within
this distance the velocity dispersion depends only weakly on the distance
to the primary for all primaries considered here. Hence, we can
meaningfully compare results from galaxies that might have different
sizes.  The estimated value of the velocity dispersion for an $L_*$
galaxy from weak lensing studies \citep{Hoeskstra, McKay01} is also
shown; the results of lensing studies are very close
to those derived from the SDSS satellite data.

The results are consistent with a slope of  $\sigma
\propto L^{0.3}$. We can write this dependence using a form analogous
to the Tully--Fisher relation:
\begin{equation}
M_B=0.02 -8.7\log(2\sigma).
\end{equation}
\noindent This relation is shown in Figure~\ref{fig:fig9}. The slope of
this relation is the same as for spiral galaxies
\citep{Verheijen}, which is shown as a long-dashed line in the figure. 
The zero point for the spiral relation was shifted down by factor of 1.6 to
match the SDSS data points; this factor is close to the naively expected
$\sqrt{3}$ correction from rotational velocity to 1D line-of-sight
velocity. The remaining difference of 8\% is difficult to assign to any
particular effect; there are several effects that could produce the small
difference (e.g., projection effects, and non-isotropic orbits). 

However, when we compute the r.m.s. velocity within a larger radius 
of 350~\kpc, we find $\sigma \propto L^{0.5}$. The reason for the steeper 
slope is a combination of the very large radius and declining $\sigma(r)$, which   will occur if the density profiles of halos at different mass are
not homologous, e.g., if the concentration of the halo decreases with
increasing mass. This is consistent with theoretical prediction.

\begin{figure}[tb!]
\plotone{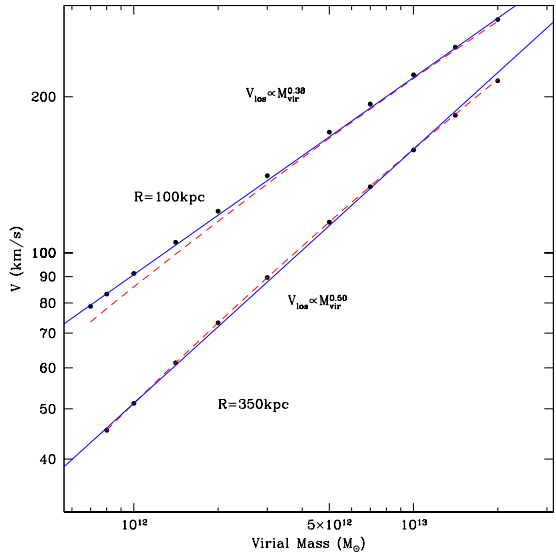}
\caption{Theoretical predictions for the line-of-sight r.m.s. velocity,
 $\sigma$, at different projected distances as the function of galaxy
 virial mass. The dots show the velocities for equilibrium NFW
 halos. The top curves and symbols are for velocities averaged for
 radii 20--100~\kpc. The bottom curves are for projected distance
 350~kpc. Power-law fits indicated in the plot with full lines provide
 very accurate approximations. The dashed curves are 1D tangential
 r.m.s. velocity at 100~kpc and at 350~kpc. }
\label{fig:fig10}
\end{figure}

It is rather straightforward to find the $\sigma$--virial mass 
relation for the NFW halos. Using the model described in
Section~\ref{ref:SecRemove} (equilibrium NFW with slightly anisotropic
velocities), we find the line-of-sight r.m.s. velocity averaged for
radii 20--100~kpc and the line-of-sight velocity for projected
distance 350~kpc. The results are shown in Figure~\ref{fig:fig10}. For
the range of masses considered here, the $\sigma$--$M$ relations are
power laws.  Just as the observed slope of the $\sigma$--$L$ relation,
the slope of the $\sigma$--$M_{\rm vir}$ relation increases with the radius
within which the {r.m.s.} velocity is calculated. For the
350~\kpc~ radius the slope 0.5 is the same as the slope of the
$\sigma$--$L$ relation for the same radius. This would imply a constant
mass-to-light ratio. This result, being formally correct, is a bit
misleading. Measured in terms of a more physically motivated quantity,   the
virial radius,  the $M/L$ ratio increases with luminosity.

Comparison with theoretical predictions of the luminosities is more
complicated and will be deferred to another paper. We use only the virial $M/L$
for two magnitudes $M_B=-20$ and $-21$.  In a recent paper, \citet{Yang}
ask which model parameters are required for the
cosmological models to explain the observed luminosity function and
the Tully--Fisher relation. The main uncertainty is the $M/L$
ratio. \citet{Yang} predict an $M/L(L)$ that is needed to account for
observations. We compare our two $M/L$ ratios with the \citet{Yang}
predictions and find them remarkably close. However, this comparison actually
required numerous corrections, including a correction for
overdensity of 180 to the virial overdensity, correction for galactic
absorption, correction for different bands, and scaling with the Hubble
constant. The corrections were provided by Rachel Somerville.
Preliminary comparison with semi-analytical models (Somerville,
private communication)
indicate a problem: observed galaxies are $\approx$0.75 mag too bright as
compared with theoretical predictions.

\section{Conclusions}

We construct samples of isolated galaxies and associated satellites using
data from the SDSS. We detect about 3000 satellites with absolute blue
magnitudes going down to $M_B = -14$.  Analysis of the satellite
velocities clearly indicates that galaxies are embedded in large halos
extending to distances up to $350~\kpc$.  Our measured velocity
dispersion for $\sim L_*$ galaxies is in agreement with measurements
from weak lensing studies.

Our analysis of galaxies selected from the SDSS database clearly
indicates that the line-of-sight r.m.s. velocity of satellites declines
with the distance to the primary galaxy. Similar results were found
for each of our three samples, suggesting that it is robust to
statistical effects and the exact definition of an isolated galaxy.
This decline agrees remarkably well with $\rho\propto r^{-3}$
predicted by all cosmological models. Both the isothermal and MOND
profiles contradict the observational results. Observations of weak
lensing from the SDSS combined with the Tully--Fisher and Fundamental
Plane relations \citep{Seljak} also finds that the galactic
halo profiles has to be steeper than isothermal at large radii.

Interlopers play an important role.  We believe that the decline in satellite
velocity was not seen previously \citep{ZaritskyWhite,
Zaritsky97, McKay02},
mainly because of either lack of statistics or insufficient removal of
interlopers. We use numerical simulations to test prescriptions for
correcting the effects of interlopers and isolation criteria. Both
effects -- interlopers and isolation criteria -- have a tendency to
overestimate of the mass of the dark matter halo and imply a velocity
profile that is too flat.

We find that the satellite velocity dispersion inside a projected
100~kpc radius and the absolute blue magnitude of the galaxy are
related as $M=0.02 -8.7\log(2\sigma)$ ($\sigma \propto L^{0.3}$).
This dependence is in agreement with the slope and the zero point of
the Tully--Fisher relation for spiral disks. If we calculate the
{r.m.s.} velocity within a radius  of $350~\kpc$, the
slope of the velocity--luminosity relation is visibly steeper: $\sigma
\propto L^{0.5}$. The likely explanation for the difference in the slopes
is an interplay between the shape of the velocity dispersion and the virial
radius of galaxies.

The mass-to-light ratio increases with luminosity as $M/L\propto
L^{0.5}$. For $M_B=-20$ (-21) we find that the virial $M/L=100$ (150).

\acknowledgments

  Funding for the creation and distribution of the SDSS Archive has been 
provided by the Alfred P. Sloan Foundation, the Participating Institutions, 
the National Aeronautics and Space Administration, the National
Science Foundation, the US Department of Energy, the Japanese 
Monbukagakusho, and the Max Planck Society. The SDSS Web site is 
http://www.sdss.org/. 

The SDSS is managed by the Astrophysical Research Consortium (ARC) for 
the participating institutions. The participating institutions are The 
University of Chicago, Fermilab, the Institute for Advanced Study, the Japan 
Participation Group, The Johns Hopkins University, Los Alamos National 
Laboratory, the Max-Planck-Institute for Astronomy (MPIA), the 
Max-Planck-Institute for Astrophysics (MPA), New Mexico State University, 
University of Pittsburgh, Princeton University, the United States Naval
Observatory, and the University of Washington.

  This research has made use of the NASA/IPAC Extragalactic Database (NED),
which is operated by the Jet Propulsion Laboratory, California Institute
of Technology, under contract with the National Aeronautics and Space
Administration.

  FP  acknowledges the hospitality of the NMSU Astronomy Department, where 
part of this work was done, and to Anatoly Klypin for financial support during
this visit. MV acknowledges the hospitality and financial
support of the MPIA, where the work was started.
MV was supported by the Los Alamos National Laboratory.
JH acknowledges the hospitality of the Instituto de Astrof\'\i sica
de Andaluc\'\i a, where some of this work was completed.

\end{document}